\documentclass[conference]{IEEEtran}
\IEEEoverridecommandlockouts
\usepackage{cite}
\usepackage{amsmath,amssymb,amsfonts}
\usepackage{algorithmic}
\usepackage{graphicx}
\graphicspath{{./figures/}}
\usepackage{textcomp}
\usepackage{xcolor}
\usepackage{glossaries}
\usepackage{url}
\usepackage{hyperref}
\usepackage{breakurl}

\newacronym{arima}{ARIMA}{Autoregressive Integrated Moving Average}
\newacronym{can}{CAN}{Controller Area Network}
\newacronym{cti}{CTI}{Cyber Threat Intelligence}
\newacronym{dbscan}{DBSCAN}{Density-Based Spatial Clustering of Applications with Noise}
\newacronym{ecu}{ECU}{Electronic Control Unit}
\newacronym{ioc}{IoC}{Indicator of Compromise}
\newacronym{IoT}{IoT}{Internet of Things}
\newacronym{lda}{LDA}{Latent Dirichlet Allocation}
\newacronym{lstm}{LSTM}{Long Short-Term Memory}
\newacronym{ml}{ML}{Machine Learning}
\newacronym{ner}{NER}{Named Entity Recognition}
\newacronym{nlp}{NLP}{Natural Language Processing}
\newacronym{nmf}{NMF}{Non-negative Matrix Factorization}
\newacronym{oscti}{OSCTI}{Open Source Cyber Threat Intelligence}
\newacronym{osd}{OSD}{Open Source Data}
\newacronym{osinf}{OSINF}{Open Source Information}
\newacronym{osint}{OSINT}{Open Source Intelligence}
\newacronym{osintv}{OSINT-V}{Validated OSINT}
\newacronym{osn}{OSN}{Online Social Network}
\newacronym{socmint}{SOCMINT}{Social Media Intelligence}
\newacronym{tfidf}{TF-IDF}{Term Frequency–Inverse Document Frequency}
\newacronym{socmati}{SOCMATI}{Social Media Automotive Threat Intelligence}
\newacronym{API}{API}{Application Programming Interface}
\newacronym{PoC}{PoC}{Proof of Concept}
\def\BibTeX{{\rm B\kern-.05em{\sc i\kern-.025em b}\kern-.08em
    T\kern-.1667em\lower.7ex\hbox{E}\kern-.125emX}}

\begin{document}
\bstctlcite{IEEEexample:BSTcontrol}

\title{Can social media shape the security of next-generation connected vehicles? \\
{% \footnotesize \textsuperscript{*}Note: Sub-titles are not captured in Xplore and should not be used
}
    \thanks{
        This study was carried out within the SERICS - Security and Rights in the CyberSpace and received funding from the European Union Next-GenerationEU (PIANO NAZIONALE DI RIPRESA E RESILIENZA (PNRR) – MISSIONE 4 COMPONENTE 2, INVESTIMENTO 1.3 – D.D. 1556 11/10/2022, PE00000014). % and within the COLTRANE-V project – funded by the Ministero dell’Università e della Ricerca – within the PRIN 2022 program (D.D.104 - 02/02/2022).
        This manuscript reflects only the authors’ views and opinions. Neither the European Union, nor the European Commission, nor the Ministry can be considered responsible for them.\\
        (*) Correspondence: stefano.dicarlo@polito.it
    }
}

% \author{
% \IEEEauthorblockN{1\textsuperscript{st} Nicola Scarano}
%     \IEEEauthorblockA{
%         % \textit{Dip. di Automatica e Informatica}\\
%         \textit{DAUIN}\\
%         \textit{Politecnico di Torino}\\
%         Turin, Italy \\
%         \href{nicola.scarano@polito.it}}
% \and
% \IEEEauthorblockN{2\textsuperscript{nd} Luca Mannella}
%     \IEEEauthorblockA{
%         % \textit{Dip. di Automatica e Informatica}\\
%         \textit{DAUIN}\\
%         \textit{Politecnico di Torino}\\
%         Turin, Italy \\
%         \href{luca.mannella@polito.it}
%       } % or 0000-0001-5738-9094} % email address or ORCID}
% \and
% \IEEEauthorblockN{3\textsuperscript{rd} Alessandro Savino}
%     \IEEEauthorblockA{
%         % \textit{Dip. di Automatica e Informatica}\\
%         \textit{DAUIN}\\
%         \textit{Politecnico di Torino}\\
%         Turin, Italy \\
%         \href{alessandro.savino@polito.it}
%     }
% \and
% \IEEEauthorblockN{4\textsuperscript{th} Stefano Di Carlo*}
%     \IEEEauthorblockA{
%         % \textit{Dip. di Automatica e Informatica}\\
%         \textit{DAUIN}\\
%         \textit{Politecnico di Torino}\\
%         Turin, Italy \\
%         \href{stefano.dicarlo@polito.it}
%     }
% }
\author{
    \IEEEauthorblockN{
        Nicola Scarano, Luca Mannella, Alessandro Savino, Stefano Di Carlo$*$,}
    \IEEEauthorblockA{
        \textit{Politecnico di Torino, Department of Control and Computer Engineering, Turin, Italy}\\
    \textit{e-mail: \{name.surname\}@polito.it}
    }
}

\IEEEoverridecommandlockouts
\IEEEpubid{\makebox[\columnwidth]{ 979-8-3503-7055-3/24/\$31.00 \copyright2024 IEEE \hfill} \hspace{\columnsep}\makebox[\columnwidth]{ }}

\maketitle

\IEEEpubidadjcol

\begin{abstract}
The increasing adoption of connectivity and electronic components in vehicles makes these systems valuable targets for attackers. While automotive vendors prioritize safety, there remains a critical need for comprehensive assessment and analysis of cyber risks.
In this context, this paper proposes a \gls{socmati} framework, specifically designed for the emerging field of automotive cybersecurity. The framework leverages advanced intelligence techniques and machine learning models to extract valuable insights from social media. Four use cases illustrate The framework's potential by demonstrating how it can significantly enhance threat assessment procedures within the automotive industry.
\end{abstract}


\glsresetall

\section{Introduction}  \label{sec: introduction}

The Automotive Industry is experiencing a substantial change in its innovation drivers, moving from traditional domains to new areas such as hardware or software. Thanks to this shift, we have experienced significant updates in various car components, including advanced driver assistance, brake systems, ignition, and engine. On the other hand, the increasing complexity of vehicular systems is opening the industry to new harms that carmakers have never experienced \cite{mandiant:2024aa}. Cyber-attacks increasingly target vehicle systems, threatening the system's security and drivers' safety~\cite{9525579,10309857,9262960}. This phenomenon is alarming carmakers and pushing innovations in automotive-oriented cybersecurity \cite{kirdi2024caracas}.

\glspl{osn} serve as platforms where many users share opinions and information across diverse topics via the Internet \cite{235461}, as they can aggregate insights and attention in response to real-world events, transcending geographical boundaries. Automatic information gathering from openly accessible sources like blogs, publications, or social networks is called \gls{osint}. This term includes all those frameworks that involve collecting and processing large amounts of data to extract knowledge from it. Specifically, \glspl{osn} over the years have become valuable data sources for researchers who have gleaned insights into various domains like political campaigns and sociological studies \cite{Aggrawal2019BehaviourOV, 10.1007/978-3-030-76228-5_37, 8954668}. 

Even though the automotive domain is experiencing an increase in cyber threats, there remains a conspicuous absence of \gls{oscti} research within the automotive domain. This paper proposes a framework for \gls{socmint} in automotive security, or \gls{socmati}, and presents four use cases that could significantly benefit from its exploitation. The system proposed will enhance the industry's capacity to anticipate, identify, and mitigate emerging threats in an increasingly interconnected and digitalized automotive landscape.

The remainder of this paper is organized as follows: Section~\ref{sec: background} reviews the current state of the art and provides the main concepts necessary to discuss the proposed approach. % and an overview of \gls{osint} within the automotive industry.
Section~\ref{sec: proposed_framework} outlines the methodologies proposed for performing \gls{socmati}.
Finally, Section~\ref{sec: conclusion} concludes the paper and discusses possible future research directions. 
%to enhance the proposed framework for automotive cybersecurity.

\section{Background and related work}  \label{sec: background}

\gls{osint} leveraging openly accessible data facilitates the retrieval of significant insights. Hwang et al. \cite{Hwang} conceptualize \gls{osint} within related terms: (i) \textit{Intelligence} refers to information of espionage, collection of data for a specific purpose, (ii) \textit{\gls{osd}} stands for unprocessed general data, (iii) \textit{\gls{osinf}} general data partially filtered on requirements, and (iv) \textit{\gls{osintv}} representing \gls{osint} with a high degree of certainty. In contrast, \gls{cti} instead refers to collecting and analyzing data related to threat data. It is a strategy widely adopted in IT systems, mainly focusing on \gls{ioc} \cite{235461}. While social media-based data collection is referred to as \gls{socmint}~\cite{Riebe2023}, etc.~\cite{threatkg}. However, the existing platform (threat sharing platform or general repositories with public \gls{oscti} reports) seldom documents and registers automotive attacks, where information remains notably scarce.

Social media platforms offer a viable resolution to the data scarcity issue, serving as prolific reservoirs of threat-related data. Platforms such as YouTube, Reddit, and Telegram are recognized as robust ecosystems harboring extensive cybercrime communities. Numerous studies \cite{8919107,9073097, 8622506} have effectively harnessed \gls{cti} derived from social network data, facilitating the extraction of advanced threat information by analyzing user interactions, communications, and preferences.

However, a few studies have investigated the potential of leveraging social media data to enhance cybersecurity in the automotive sector. Oberti et al. \cite{10207148} focuses solely on Twitter as a data source, presenting a comparable workflow but with limited detail on data processing techniques and methodologies. Additionally, despite providing a comprehensive overview of standards and context in the automotive domain, there appears to be a lack of understanding of the broader \gls{oscti} field, encompassing cybersecurity and data mining. Instead, the work of Bertoglio et al. \cite{BertoglioPensoSenniGuidottiMagnani} suffers from inaccuracies in \gls{cti} terminology, a lack of detailed methodological discussion spanning data sources to word embedding and \gls{ml} techniques, and vague definitions of solutions, with inadequate perspectives on potential applications and future work.

\begin{figure}[ht]
    \centering
    \includegraphics[width=0.99\linewidth]{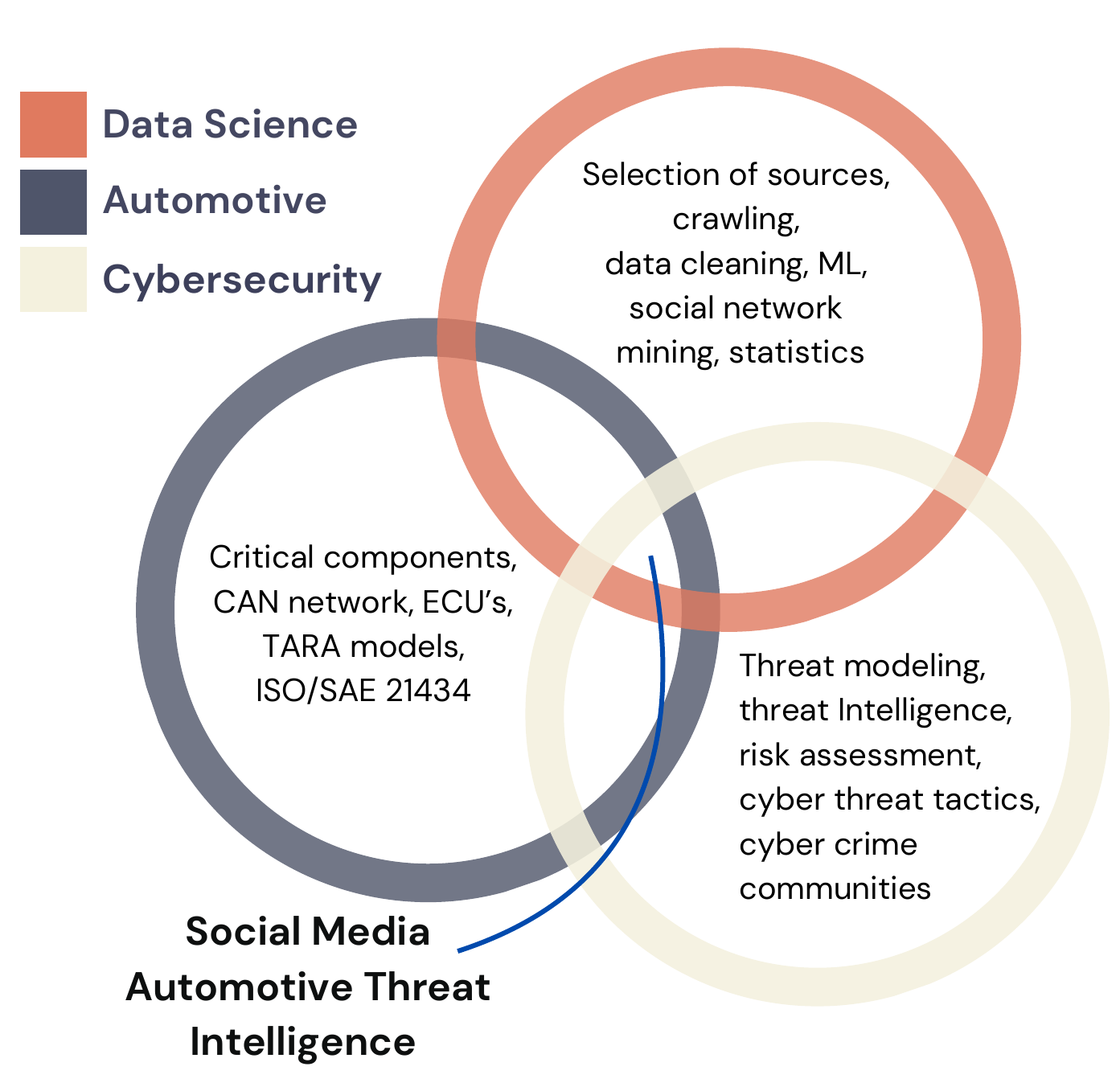}
    \caption{This figure highlights three major competencies to design and implement the proposed framework.}
    \label{fig:Venn}
\end{figure}

In this context, supporting \gls{socmati} must complement competencies across all the three macro areas interested: automotive, data science, and cybersecurity \ref{fig:Venn}.

For this reason, we propose a new framework that exploits the work done in the existing \gls{oscti} field and adapts it to the needs and requirements of the growing automotive security domain.

\section{Proposed framework}  \label{sec: proposed_framework}
\begin{figure*}[tb]
    \centering
    \includegraphics[width=0.99\textwidth]{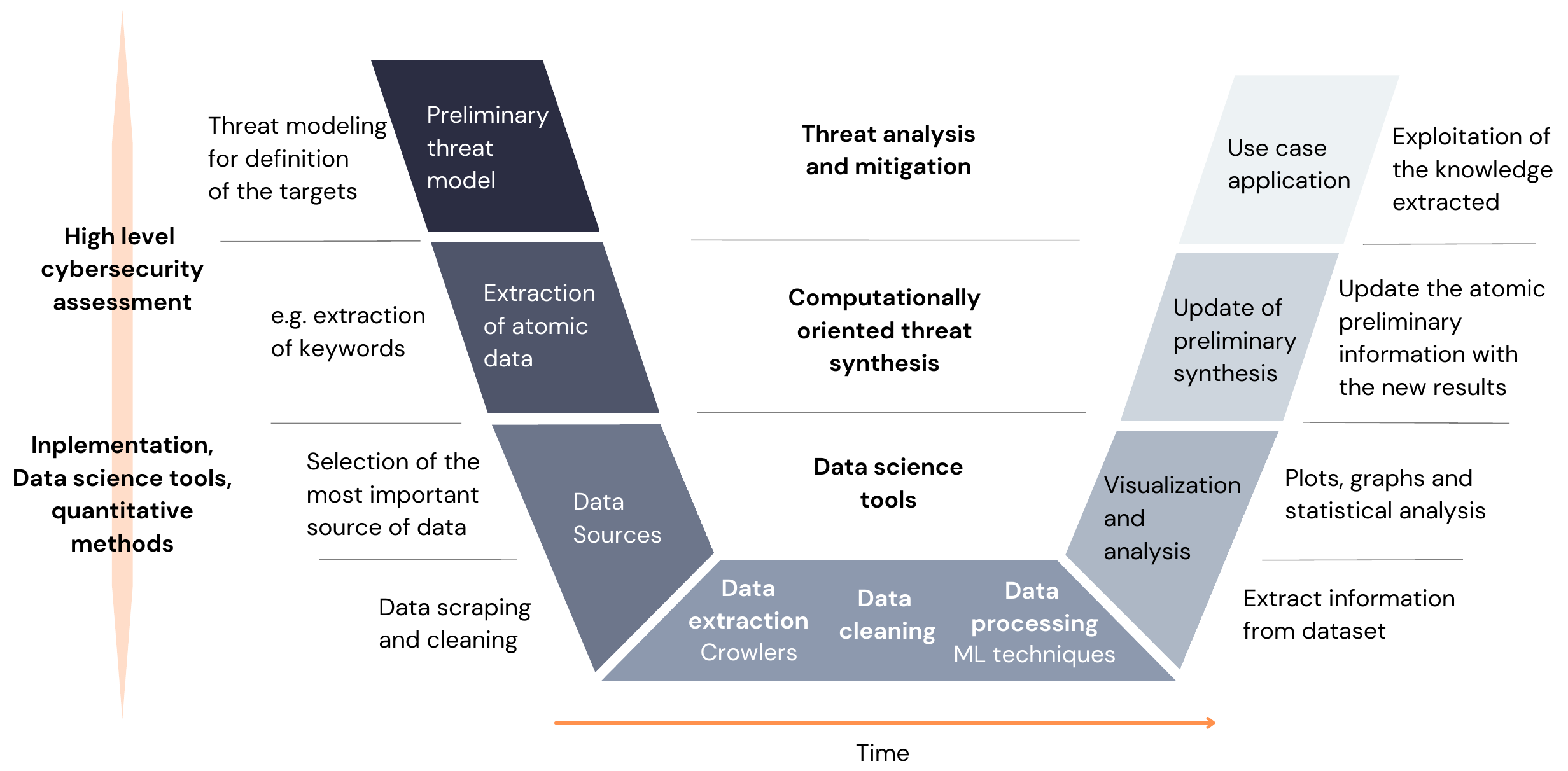}
    \caption{The picture illustrates the V-model of the framework flow presented in the paper. The upper part of the model represents the initial and final stages of the workflow, focusing on structuring and validating the extracted threat information, such as the threat model. In contrast, the lower part of the model details the implementation steps. It begins with extracting preliminary data from the qualitative threat analysis, followed by a data science pipeline involving data extraction, cleaning, and processing. It concludes with visualizations and updates to the preliminary synthesis.}
    \label{fig: V_model}
\end{figure*}

The framework integrates \gls{socmint} into a cyber threat mitigation strategy. In particular, the data-driven approach proposed using a corpus of open data allows the automation of part of the cyber security assessment and improves its accuracy, depth, and speed. The approach proposed at this stage requires the cybersecurity expert to supervise all the phases of the data-driven approach. 
The framework can be modeled as a pipeline of seven phases as illustrated in Figure~\ref{fig: V_model} following a V-model. The initial phase involves collecting information on vulnerabilities and potential threats, thus building a preliminary threat model. The subsequent phase focuses on defining and extracting atomic data like keywords, which is the starting point for data extraction and analysis methods. We identify and select open-source data in the third phase, concentrating primarily on social media platforms. This involves thoroughly evaluating various social media channels, groups, and chats to identify those pertinent to our research objectives. The next phase entails the application of a data science pipeline to the extracted data, leveraging \gls{ml} techniques. The final phases, on the right side of the V-model, focus on exploiting the extracted knowledge. These include analyzing and visualizing the results, updating the preliminary synthesis with new findings, and exploiting the processed understanding on a specific use case.

This systematic approach ensures that the framework captures critical threat information and repeats the process multiple times, continually enhancing its threat model through iterative updates and refinements. In the following sections, the paper briefly describes the key components of our framework.

\subsection{Data acquisition and cleaning}
\label{subsec: data-acquisition}

The proposed framework relies extensively on data, its most valuable and sensitive component. Therefore, identifying reliable data sources and implementing rigorous data cleaning, transformation, and processing pipelines are fundamental to the framework's effectiveness and reliability.
Communication platforms such as Telegram, Reddit, and YouTube have emerged as significant data reservoirs for cybersecurity analysis due to their substantial engagement from cybercrime communities. Although X (formerly known as Twitter) has restricted open access to its \gls{API} by imposing a fee, other publicly accessible social networks continue to offer invaluable data that can be leveraged for comprehensive cybersecurity research. Conversely, Meta's platforms require specific ethical safeguards (ethical committee agreement) for automatically accessing third-party data since they might present user's personal information.
These platforms provide official \glspl{API} that allow researchers to gather text-based data and pertinent metadata methodically. In particular, we can categorize the data scraped into three key types:
\begin{itemize}
    \item Textual Data: posts, comments, discussions, video transcripts, chat logs.
    \item Interaction Information: metrics include replies, likes/dislikes, shares/retweets.
    \item Metadata: timestamps, geographical locations (when available), user profiles.
\end{itemize} 

Once the data is collected, the pre-processing phase is crucial for ensuring its cleanliness and standardization, which is essential for effective analysis. The processed data is then meticulously organized into structured datasets.

\subsection{Processing techniques}
\label{subsec: data-processing}

The framework relies on a \gls{ml} pipeline to extract detailed information from the corpus of data crawled from network sources. Based on these data types presented in Section \ref{subsec: data-acquisition}, we have selected a list of processing techniques for social media analysis and threat intelligence that have been successfully deployed.

\textbf{Time Series Analysis} is a specialized method for examining a sequence of data points collected over regular intervals. In this type of analysis, data points are consistently recorded at set intervals throughout a specified duration instead of being captured intermittently or randomly. Promising are the \gls{arima} models used for short-term forecasting of trends in threat events and attack patterns activity and the \gls{lstm} networks for long-term trend analysis that allow capturing seasonal and temporal effects over extensive datasets instead.

\textbf{\gls{nlp} methods} combine the techniques of linguistics and computer science to create algorithms that can interpret human language. They can be helpful in processing and extracting information from comprehensive corpora of text data. In particular, the applications of \gls{socmint} and \gls{cti} could benefit from exploiting methods to extract keywords related to a specific topic. In this context, word embedding methods (\gls{tfidf}, Word2Vec, etc) and topic modeling techniques (e.g., \gls{ner}, \gls{lda}, \gls{dbscan}, etc) offer great potential.

\textbf{Interaction Analysis} exploits network analysis for text mining. Analyzing large corpora of data from social media can unveil connections between concepts, entities, or documents by representing them as nodes in a network and their relationships as edges. It also allows us to identify clusters or communities within a network, revealing cyber-criminal groups and threat topics and enhancing our understanding of the underlying structure of the data.

\section{Use cases}  \label{sec: possible_applications}
In this section, we present four use cases that may deeply benefit from exploiting the social media threat intelligence approach described in this paper. 
% For each of these use cases, we briefly describe how our framework can be used to enhance the security of the specific scenario.

\subsubsection{Learn attack tactics information}
Attackers use social network crime communities to exchange information about attack strategies. The ability to analyze chats, groups, blogs, or videos in an automated way can allow us to learn new tactics or update old ones. The automatic update allows carmakers and security experts to adopt or update new defense mechanisms. In this case, \gls{nlp} techniques are key. In particular, it is required to detect topics inside the whole corpus of data and, after selecting the topic, extract hot words inside each single cluster. Then, using anomaly detection techniques could allow us to detect new words related to new techniques or technologies used for a specific attack.

\subsubsection{Threats trend detection and prediction}
The collected data can be used to detect and predict cyber threat trends. Trends can be analyzed based on geographical data, making these predictions granular and specific to different regions. The performance of this type of analysis requires \gls{nlp} capabilities alongside time series analysis methods. In this case, extracting quantitative metrics is critical to perform trend prediction. Examples of valuable data in this context are engagement and interaction metrics extracted from content on a specific topic.

\subsubsection{Define threat intelligence metrics}
Defining metrics to assess and extract relevant data is crucial in evaluating the threat intelligence process. Inspired by \cite{235461}, which presents a set of metrics for comparing threat intelligence sources and assessing their suitability for specific purposes, this approach aims to ensure a thorough assessment of threat intelligence effectiveness. Metrics can be extracted to assess a channel's quality or the relevance of a specific search query in a YouTube search. The formal definition of metrics allows researchers to define better the extent to which the data collected can meaningfully support their intended uses.

\subsubsection{Data Driven Risk Assessment}
Risk assessment is a well-known mitigation strategy, already described in automotive ISO/SAE 21434~\cite{isosae}. However, this is usually done by directly analyzing the platform to protect it. Introducing \gls{socmati} can help extract risk metrics and compute risk models, contributing to cybersecurity and safety. This use case needs the construction of merging protocols that exploit the information extracted using \gls{socmint} with already existing risk models. A \gls{PoC} of the use of extracted Twitter data for automatic update of the risk model is presented in \cite{10207148}.
\section{Conclusion and future work}  \label{sec: conclusion}
This paper proposed a framework that exploits \gls{cti} techniques for building a social media cybersecurity framework tailored to the automotive domain: \gls{socmati}. The proposed framework leverages intelligence methods and data analysis techniques to extract valuable information from social media. Several use cases demonstrate the framework's potential to enhance threat assessment procedures in the automotive industry significantly.

Future research will focus on implementing and evaluating this framework in real-world scenarios. This initial development represents only the beginning; numerous challenges remain in elevating automotive cybersecurity to match the expertise found in traditional IT sectors. A particularly urgent task is the development of ontologies to improve the understanding of attack patterns and response strategies.

In conclusion, the diffusion of \gls{oscti} in the automotive industry is essential. It will enable the sector to anticipate better, identify, and mitigate emerging threats in an increasingly interconnected and digitalized automotive landscape.

\vspace{12pt}

\color{black}

\bibliographystyle{IEEEtran}
\bibliography{IEEEabrv,biblio}

\end{document}